\documentclass[fleqn,10pt]{wlscirep}
\usepackage{graphicx}
\usepackage{caption}
\usepackage{subcaption}
\usepackage{bbm}
\usepackage{amssymb}

\title{Higher-dimensional performance of port-based teleportation}
\author[1,*]{Zhi-Wei Wang}
\author[2,+]{Samuel L. Braunstein}
\affil[1]{Tang Aoqing Honors Program in Science,  College of Physics,
Jilin University, Changchun, 130012, People's Republic of China}
\affil[2]{Computer Science and York Centre for Quantum Technologies,
University of York, York YO10 5GH, United Kingdom}
\affil[*]{zhiweiwang.phy@gmail.com}
\affil[+]{sam.braunstein@york.ac.uk}

\begin{abstract}
Port-based teleportation (PBT) is a variation of regular quantum
teleportation that operates without a final unitary correction. However,
its behavior for higher-dimensional systems has been hard to calculate
explicitly beyond dimension $d=2$. Indeed, relying on conventional
Hilbert-space representations entails an exponential overhead with
increasing dimension. Some general upper and lower bounds for various
success measures, such as (entanglement) fidelity, are known, but
some become trivial in higher dimensions. Here we construct a
graph-theoretic algebra (a subset of Temperley-Lieb algebra) which
allows us to explicitly compute the higher-dimensional performance of
PBT for so-called ``pretty-good measurements'' with negligible
representational overhead. This graphical algebra allows us to explicitly
compute the success probability to distinguish the different outcomes
and fidelity for arbitrary dimension
$d$ and low number of ports $N$, obtaining in addition a simple upper
bound. The results for low $N$ and arbitrary $d$ show that the fidelity
asymptotically approaches ${N}/{d^2}$ for large $d$, confirming the
performance of one lower bound from the literature.
\end{abstract}
\begin{document}

\flushbottom
\maketitle
\thispagestyle{empty}

\section*{Introduction}

Port-based teleportation (PBT) \cite{ishizaka2008asymptotic} is a
variation of the conventional quantum teleportation, which can be used
as a universal processor. In PBT, Alice (the sender) and Bob (the
receiver) share $N$ pairs of entangled quantum states (without loss of
generality we assume these to be maximally entangled). Then Alice
performs a joint measurement (a positive operator valued measurement,
POVM, $\{\Pi_{AC}^i\}$) on her resource and the input quantum state
$\sigma^{\text{in}}_C$ that she wishes to teleport. Finally, Alice tells
Bob the measurement outcome $i\in \{0,1,\ldots,N\}$. If $i=0$ the
teleportation is considered to have failed, otherwise, Alice's input
state $\sigma_C^{\text{in}}$ will have been teleported to the $i$th
entangled partner (or port) in Bob's possession.
Unlike conventional teleportation
\cite{bennett1993teleporting}, PBT does not require any corrective unitary
operation at Bob's side other than to discard the unused ports.


\begin{figure}[ht]
\centering
\includegraphics[width=0.5\textwidth]{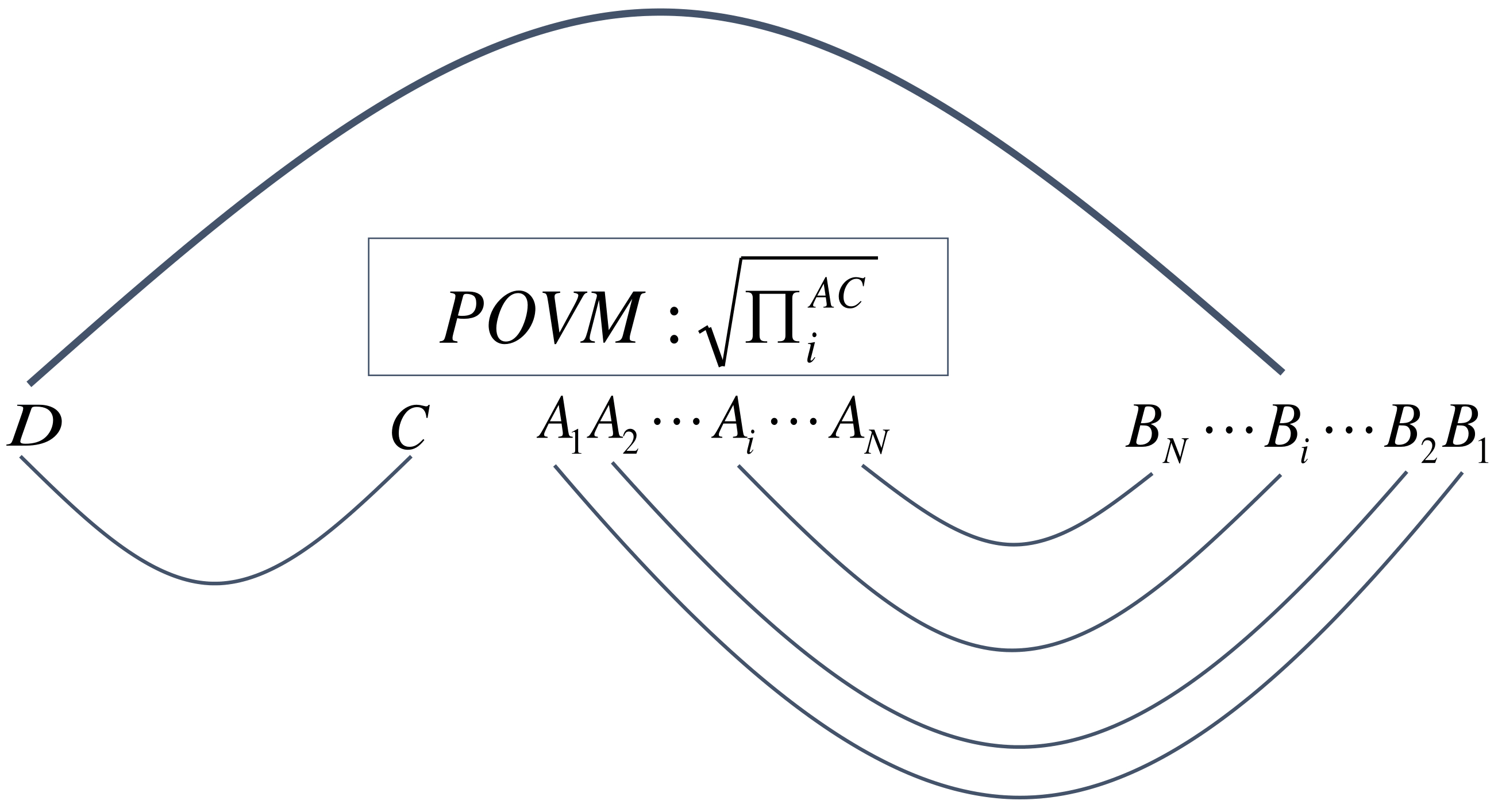}
\caption{\textbf{PBT protocol for teleporting entanglement.} When the
input system $C$ is half of an entangled state $\sigma_{DC}$, PBT will
`swap' the entanglement so that now systems $D$ and $B_i$ (corresponding
to port $i$ in Bob's possession) become entangled. The joint quantum
state $\sigma_{DC}$ will have been transferred to $\sigma_{DB_i}$.
Finally, if $\sigma_{DC}$ is initially maximally entangled, then the
`global' fidelity (including the entanglement with system $D$) of this
teleportation is given by the so-called entanglement fidelity for this
process.}
\label{P0}
\end{figure}

The lack of a final correcting unitary for the PBT protocol has led to a
number of important theoretical implications, including universal
quantum processing, non-local processing, etc. (for a more complete
discussion of applications see
ref.~\cite{ishizaka2015some,pirandola2015advances}). Despite its clear
theoretical importance the achievable performance of PBT is largely
restricted to low dimensional systems \cite{ishizaka2008asymptotic}
($d=2$ with arbitrary $N$, and $d=3$ for $N\le 6$). There are several
papers giving bounds to the performance of PBT more generally
\cite{ishizaka2015some,beigi2011simplified,
ishizaka2009quantum,pitalua2013deduction}, but they do not explicitly
calculate the achievable fidelity (or other performance measures) for
higher dimensional systems. In addition, some bounds become trivial in
higher dimensions, for example some lower bounds become negative for
higher dimensional systems \cite{ishizaka2015some,beigi2011simplified,
ishizaka2009quantum,pitalua2013deduction}. Thus it is not clear that we
may rely on these bounds for evaluating the theoretical performance of
PBT in its manifold applications.

Since a $d$-dimensional quantum system can be written in terms of
$\log_2 d$ qubits, we might naively expect that the PBT of a
$d$-dimensional system would be equivalent to $\log_2 d$ repetitions of
the PBT protocol on this many qubits. This reasoning is incorrect for at
least two reasons. First, PBT does not operate with either unit success
probability or with unit fidelity, so this decomposition would not yield
the same performance except possibly in the asymptotic limit of an
infinite number of ports. Second, in the most general scenario, the $N$
ports at Bob's side could be non-locally distributed among $N$ Bob's
(Bob$_1$, Bob$_2$, etc.). In this scenario, a single PBT would transfer
a single $d$-dimensional system to a single (rnadom) Bob; however, the
repeated protocol on $\log_2 d$ qubits would transfer them across a
random distribution of non-locally separated Bobs. The achievable
performance of PBT for higher-dimensional quantum system is therefore
interesting in its own right.

Here we construct a graphical algebra for computing the explicit
performance of PBT based on so-called ``pretty-good-measurements''. The
graphical representations thus produced have a size which is independent
of the Hilbert-space dimension $d$, but is instead related only to the
number of ports $N$. This allows us to compute the success probability
to distinguish the different outcomes and also the fidelity for PBT even
for high dimensions, though currently our analysis is limited to small
$N$ (we expect that by identifying those graphs with the largest
contributions to performance, that we could in future determine the
achievable behavior for arbitrary $d$ and $N$). The graphical algebra we
use is a subalgebra of the Temperley-Lieb algebra and we discuss some of
the properties of this subalgebra at the end of the article.

\section*{Results}

\begin{figure}[!ht]
\centering
\includegraphics[width=0.5\textwidth]{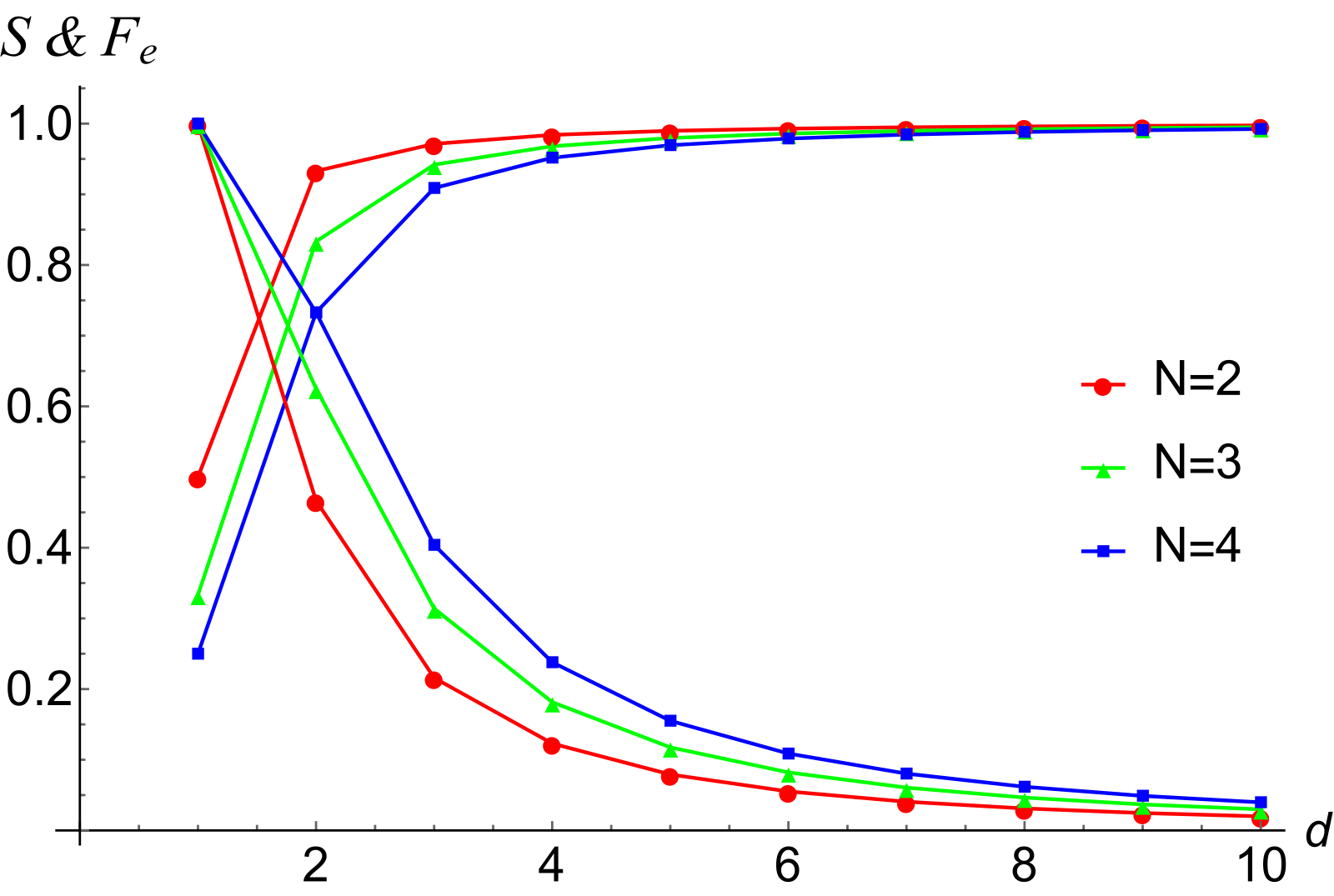}
\caption{\textbf{The performance of PBT for N=2,3,4.} Here we plot the
success probability $S$ and entanglement fidelity $F_e$ of PBT for
$N=2,3,4$ as a function of dimension $d$. As $d$ increases, the success
probability for the POVM approaches one, for increasing $N$ we find
this success probability decreases.}
\label{SF}
\end{figure}

With our graphical algebra, we have calculated the success probability,
$S$, and fidelity of PBT for $N=2,3,4$ for arbitrary $d$. We give
explicit expressions of the success probability in Eq.(\ref{success
probability}) and one can obtain the entanglement fidelity, $F_e$, and
the average fidelity, $F$, very easy through their relations
\cite{ishizaka2008asymptotic}
$F_e={N\,S}/{d^2}$ and $F=(d\,F_e+1)/(d+1)$.
\begin{equation}
\begin{split}
&S_{(N=2)}= \frac{d+\sqrt{d^2-1}}{2d} \\
&S_{(N=3)}= \frac{5d^2+d\left (2\sqrt{d-2}\sqrt{d+1}
+2\sqrt{d+2}\sqrt{d-1} \right )-2\sqrt{d-2}\sqrt{d+1}
+2\sqrt{d+2}\sqrt{d-1}-2 }{9d^2} \\
&S_{(N=4)}= \frac{1}{16d^3}\left[\right.7d^3
+d^2\left (2\sqrt{d^2-2d}+2\sqrt{d^2+2d}
+\sqrt{d-3}\sqrt{d+1}+\sqrt{d+3}\sqrt{d-1}
+3\sqrt{d^2-4}\right )- 3d\left(\right.1 +\\
&\sqrt{d-3}\sqrt{d+1}+\sqrt{d+3}\sqrt{d-1}\left.\right)
-2\sqrt{d^2-2d}-2\sqrt{d^2+2d}+2\sqrt{d-3}\sqrt{d+1}
+2\sqrt{d+3}\sqrt{d-1}-3\sqrt{d^2-4}\left.\right]
\end{split}
\label{success probability}
\end{equation}
We plot these measures of success probability in Fig.~\ref{SF}. Although
only calculated for small $N$, we find that: (i) $S$ quickly approaches
one as $d$ increases; (ii) the entanglement fidelity $F_e$ goes to zero
asymptotically with $d$ with roughly the exponent of $-2$; and (iii) the
average fidelity $F$ is only slightly higher than $F_e$ (note that $F\ge
F_e$ must always hold). Across common values of $N$ and $d$ these
results agree with the original analysis of PBT
\cite{ishizaka2008asymptotic}.

\begin{figure}[!ht]
\centering
\includegraphics[width=0.8\textwidth]{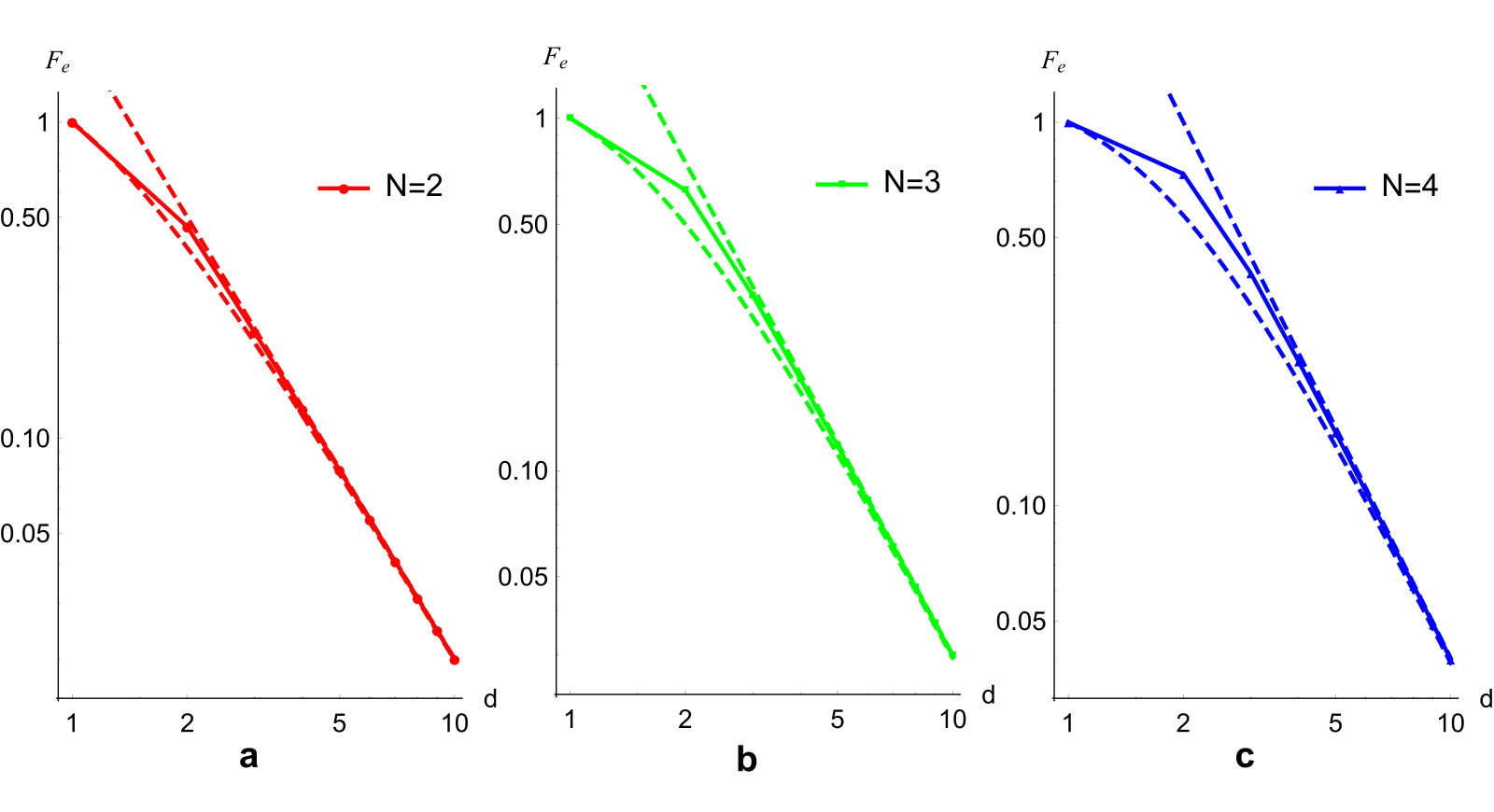}
\caption{\textbf{The entanglement fidelity and its upper and lower
bounds of PBT for small N.} Here we give a log-log plot of $F_e$ versus
$d$ plot for $N=2,3,4$ computed using our graphical-algebraic
techniques. The analytic form for the upper and lower bounds are given
in the text. For $d^2\gg N$ increases, the lower bound asymptotes
towards the straight-line upper bound with gradient $-2$ in this log-log
plot.}
\label{N234}
\end{figure}

We give a trivial upper bond to the entanglement fidelity $F_e\leq
F_e^\text{up}\equiv N/d^2$ because the success probability is itself
upper-bounded by one. A lower bound has been derived
\cite{ishizaka2015some} to the entanglement fidelity is $F_e\ge
F_e^\text{low}\equiv {N}/(d^2+N-1)$. Figure~\ref{N234} demonstrates that
the lower bound corresponds to a good approximation for lower $N$ and
our trivial upper bound appears to be reached asymptotically for higher
dimensions. Note that for a general number of ports satisfying $d\gg N$,
we have $F_e^\text{up}\approx F_e^\text{low}$; and hence the actual
entanglement fidelity for PBT is tightly constrained in these
circumstances for asymptotically large dimensionalities.


\section*{Discussion}

Our graphical algebra allows us to explicitly determine the
higher-dimensional performance of PBT. However, as $N$ increases, the
size of the algebra will exponentially increases. We tame this growth
substantially by relying on the underlying permutation symmetry amongst
the ports which allows us to only consider those diagrams within a
``conjugacy class''. So far, we have not been able to fully utilize this
symmetry in the final steps of the computation of the success
probability where the contribution of every member of a conjugacy class
needs to be separately evaluated. This difficulty has limited our
ability to extend our results beyond $N=4$, though for arbitrary $d$. It
is our expectation that further simplifications can be found to allow
our approach to be used for higher $N$; it may even be that the
asymptotic behavior can be determined from just the study of a limited
number of conjugacy classes. Such extension will be the subject of
future research.

Finally, the properties of the algebra are not yet completely
understood. The pretty-good-measurements are determined by a positive
operator $\rho$ whose eigenvalues are of the form $d\pm i$ with $i< N$.
Naively, for $d<N$ extra negative eigenvalues would seem to appear in
the general expressions constraining $\rho$. In fact, these strictly
vanish for any specific $d<N$ and such terms exactly cancel in the final
results for sucess probbility, etc. (see for example Eq.~(\ref{success
probability})). For these lower dimensionalities this affect seems
to be related to singular-perturbation theory, though here with a
geometric interpretation where singular behavior correspond to
subspaces which vanish from the problem as $d$ is successively decreased.
This geometric interpretation of singular-perturbation theory may be
new and if so deserves further examination.

\section*{Methods}

\subsection*{Mathematical representation of entanglement fidelity as a
graphical algebra}

If the teleportation is successful, Bob will discard all of his ports
except the $i$th port $B_i$ which ideally will become the unknown
quantum state $\sigma^{in}_C$. According to the original article on
PBT\cite{ishizaka2008asymptotic}, we may obtain a mathematical
expression for the entanglement fidelity for PBT as
$F_e=\sum_{i=1}^N {\text tr}_{AC}\left [\Pi_{AC}^i
\sigma _{AC}^{i}\right]/d^2$, where $\{\Pi^i\}$ are the set of POVMs,
$\sigma^i_{AC}\equiv\left(|{\Phi_{A_i C}}\rangle
\langle{\Phi_{A_i C}}|
\otimes{\mathbbm{1}_{\bar{A_i}}}\right)/d^{N-1} and$
$| {\Phi}\rangle\equiv\sum_{k=0}^{d-1}|{k k}\rangle/\sqrt{d}$ is
the (canonical) maximally entangled state (here $\bar{A_i}$ means
every system except $A_i$). Details of the derivation are in
Supplementary Derivation S1.

According to the above expression for $F_e$, the key part needed to
determine the entanglement fidelity of PBT is to calculate the trace of
$\Pi_{AC}^i\sigma^{i}_{AC}$. For the choice of so-called
``pretty-good measurements'' \cite{ishizaka2008asymptotic},
$\Pi_{AC}^i\equiv\rho^{-1/2}\sigma^{i}_{AC}\rho^{-1/2}$ where
$\rho=\sum_{i=1}^{N}\sigma^i_{AC}$. Therefore, the
key element of the calculation comes down to analysing the properties of
$\sigma^i_{AC}$. For convenience, we drop the system labelling subscripts
and shall use $\sigma_i$ to denote $\sigma^i_{AC}$, etc.,
in the following sections.

\begin{figure}[!htbp]
\centering
\includegraphics[width=0.5\textwidth]{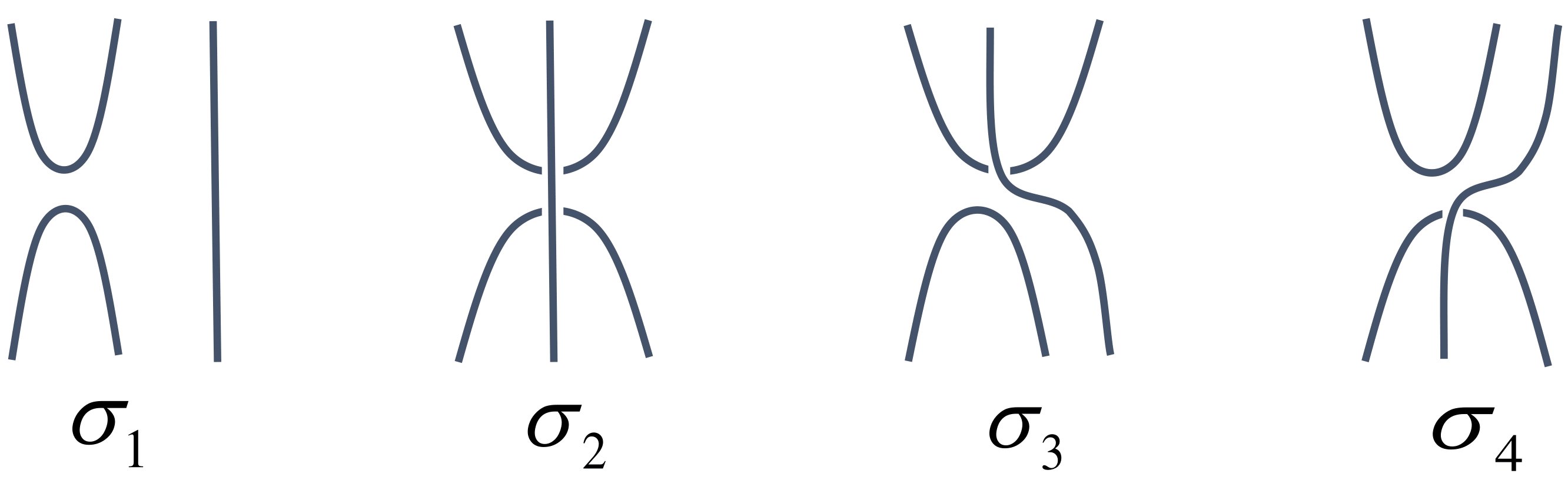}
\caption{\textbf{The graph algebra for N=2.} When $N=2$, we may write
down a `graphical basis' for analyzing PBT which we denote for simplicity
here as $\sigma_1$, $\sigma_2$, $\sigma_3$, $\sigma_4$. Note that
to describe $\rho$ we need only the first two of these graphs, however,
to describe arbitrary powers of $\rho$ we will in general
require all of these graphs.}
\label{P1}
\end{figure}

Inspired by Fig.~\ref{P0}, we constructed a graphical
algebra to represent $\sigma_i$ in a manner independent of $d$. Let us
explain the graphical algebra for the case of two ports, when $N=2$.
The shape $\bigcup$ will denote the (unnormalized) ket $|\Phi\rangle$;
the shape $\bigcap$ will denote the (unnormalized) ket $\langle\Phi |$,
a verticle line will denote the identity operator.
So, in Fig.~\ref{P1}, $\sigma_1$ denotes the
operator $|jji\rangle\langle kki|
=|jj\rangle\langle kk|\otimes|i\rangle\langle i|$ (where there is an
implied sum over the repeated indices). Thus, the graph denoted
$\sigma_1$
corresponds to the (unnormalized) maximally entangled state on subsystems
0 and 1 and the maximally mixed state on subsystem 2. The left-most subsystem
in our graphs is labelled zero and permutations of it do not contribute
to the algebra. Were we to `normalize' this graph, we could multiply it by $1/d^N$.

\begin{figure}[!htbp]
\centering
\includegraphics[width=0.8\textwidth]{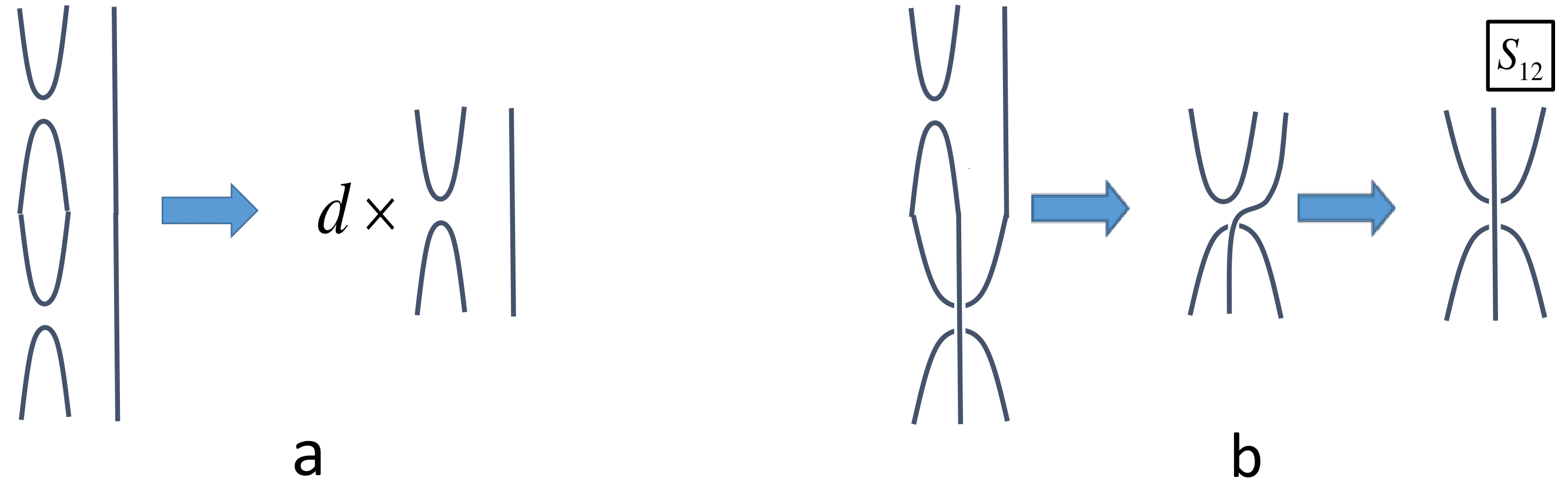}
\caption{\textbf{The multiplication rule when N=2.} a) Here we
illustrate the multiplication $\sigma_1 \sigma_1$. Loops reduce to a
numerical factor $d$; b) Here we illustrate the multiplication $\sigma_1
\sigma_2$. Lines may be internally rearranged so long as the ordering
(or labelling) of all line ends is unchanged. Note that standard forms
can be preduced by at most the pre- or post-applicaiton of permutations
among the ports (i.e., excluding subsystem zero). The result of the
multiplication is found to be labelled as $\sigma_4$ in Fig.~\ref{P1}.
This in turn may be written in terms of $\sigma_2$ as $S_{12}\sigma_2$
(here $S_{12}$ is a simple permutation operator which swaps ports $1$
and $2$).}
\label{P2}
\end{figure}

Since the graphs represent operators, they may be added and multiplied
like operators. Specifically, to multiply two operators denoted by
graphs, the graph for the leftmost operator is placed above that of the
rightmost operator and the lines of their respective subsystems are
joined in the natural manner (see Fig.~\ref{P2}). Any loop reduces to a
trace of the identity operator on a $d$-dimensional subsystem, i.e., the
factor $d$. Other simplifications come from internally rearranging the
lines. Finally, the resultant operators can always be placed into a
standard form be at most pre- and post-muliplication by permutation
operators on the ports (i.e., those subsystems in Bob's hand labelled
$1,\ldots,N$). Some of these rules are shown in Fig.~\ref{P2} for $N=2$.

The multiplication rules can be summarized by the result
\[ \sigma_i\sigma_j =
\begin{cases}
d\,\sigma_{i}                         & \quad \text{if } i=j\\
S_{ij}\,\sigma_{j} = \sigma_j\,S_{ji}  & \quad \text{if } i\neq j,
\end{cases}
\]
all for $i,j\in\{1,2\}$. All other multiplications may be obtained by
noting that $S_{12}S_{12}$ reduces to the the trivial (identity)
permutation.

With the above results, we may graphically evaluate the success
probability as $S=\frac{1}{2}(1+\sqrt{1-d^{-2}})$. The details of the
derivation are given in Supplementary Derivation S2. For higher $N$ we
must allow for an accumulation of swap operations on any pair of ports.
When combined, these swaps form arbitrary permutations as we shall now
consider.

\subsection*{Graphical Algebra for $N>2$}

Although the result for $N=2$ is very simple, when we use our approach
for larger $N$ the size of the algebra increases very quickly. In order
to compute the success probability using pretty-good-measurements we can
restrict ourselves to fully permutation symmetric expressions. For
example the quantity $\rho\equiv \sum_{i=1}^N \sigma_i$ appears in the
form $\rho^{-1/2}$. By relying on permutation symmetry we may eliminate
from active consideration any pre-applied permutation of our graphs.
However, this still leaves with an effectively exponential number of
graphs differing by a post-applied permutation.

Our main strategy relies on the fact that the algebra generated solely
from $\rho$ (consisting of the set of elements formed from the series
$\rho$, $\rho^2$, $\rho^3$, $\ldots$) must eventually close for any
finite $N$. Because of this we will be able to express any power of
$\rho$ (such as $\rho^{-1/2}$) as a finite order polynomial in $\rho$
itself. Further, beause every element in this algebra commutes with
every other element, we may write out the minimal `polynomial of
closure' directly in the basis where $\rho$ is diagonal. In this way,
this polynomial automatically correspond to a polynomial whose solutions
correspond to the eigenvalues of $\rho$. Note that given these
eigenvalues we can immediately represent any power of $\rho$, and in
particular $\rho^{-1/2}$, as a polynomial of this algebra generated from
$\rho$ itself. See Supplementary Derivation S3 for proofs for these
statements.

\begin{figure}[!ht]
\centering
\includegraphics[width=0.5\textwidth]{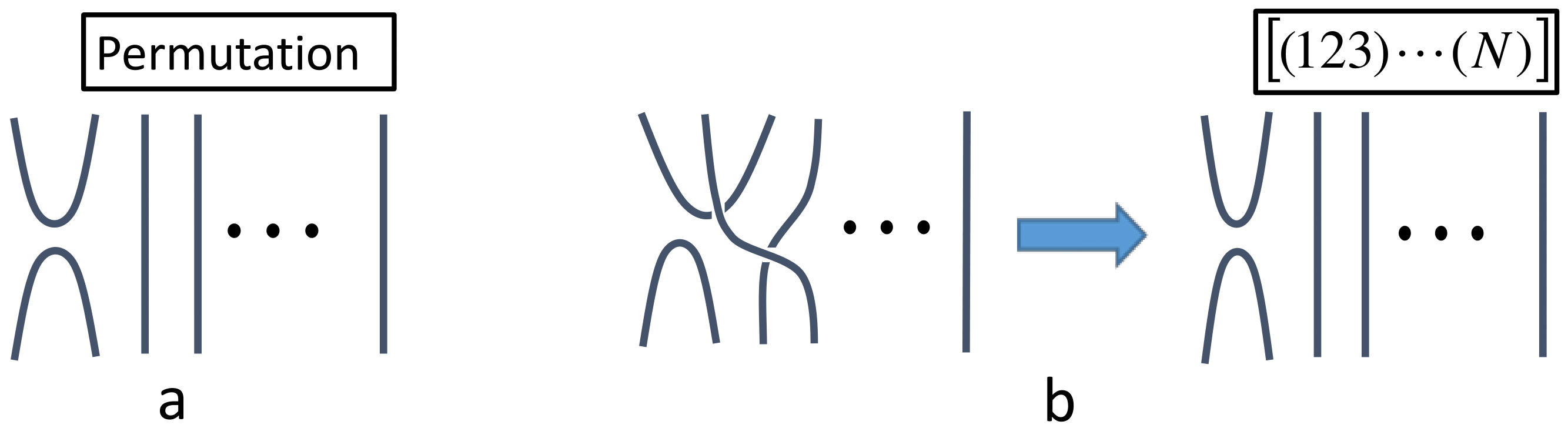}
\caption{\textbf{The graphical algebra with permutations for larger N.}
a) We express permutations by the conventional `cycle' notation of the
permutation group. Taking into account the global permutation symmetry
every operator in our algebra can be summarized by a
sum over post-permutations on $\sigma_1$. b) Shows an example expressed
as a simple three-cycle
$\left[(123)\dotsc(N)\right]\,\sigma_1$. }
\label{P3}
\end{figure}

Computing the closure polynomial involves (sums over) simple
repeated multiplications of terms like $\sigma_i \sigma_j$. Each
such multiplication will reduce to a generic term like $S_{ij} \sigma_j$.
Iterating this leads to a sequence of post-swap operations, or equivalently
a single post-permutation operation. This means that a typical term
from $\rho^n$ will consist of a sum over terms shown in Fig.~\ref{P3}.
We denote permutations using the conventional cyclic notation,
so that $(123) ...(N)$ denotes shifting $1\rightarrow 2$, $2\rightarrow 3$
and $3\rightarrow 1$, etc. To represent this as a unitary operator permuting
the ports we surround it by square brackets, e.g.,
$\left[(123) ...(N)\right]$.

To further reduce the complexity of our results we can further rely on
global permutation symmetry to ensure that once one term with a specific
`class' of permutations appears, e.g., a single three cycle such as for
the permutation $(123)(4)\cdots(N)$, then every three-cycle permutation
up to port relabelling will also appear (equivalence under relabelling
yields an equivalence class of permutations). By separately being able
to compute the size of such conjugacy classes we need only explicitly
store the appearance of an individual exemplar from each class when it
appears. The computation of the size of these conjugacy classes is given
explicitly in the next section.

The polynomial expressions for $\rho^{-1/2}$ involve coefficients
which explicitly depend on $d$. Finally, we can use our graphical
representation to simplify and explicitly compute the success
probability which is given by
${\rm tr}(\rho^{-1/2}\sigma_i\rho^{-1/2}\sigma_i)$.
Computing the trace is easy. The terms are mutiplied, and put into a
sum over the standard form $\left[{\rm Permutation}\right] \sigma_1$;
for each graph we close the join the $i$th line at the top with the same
line at the bottom and then count loops --- each loop yielding a factor
of $d$. For example
${\rm tr}\left\{\left[(12)(3)(4)\right]\sigma_i\right\}=d^3$.
This allows us to directly compute the success probability and as already
mentioned the entanglement fidelity and average fidelity then
immediately follow. All calculations we performed using Mathematica.

\subsection*{Conjugacy Classes}

The number of the conjugacy classes for different $N$ is very useful
because it provides an upper bound on the number of distinct eigenvalues
of the operator $\rho$ and can tell us how large the algebra can be. We
find the the number of the conjugacy classes can be determined by a
variation of integer partition theory.

Firstly, let us review conventional integer partition theory. A
partition of a positive integer $n$ is a way of writing $n$ as a sum of
positive integers witout regard to the order in which the sum is written
out \cite{rademacher1938partition}. The number of partitions of an integer
$n$ is given by the so-called partition function $P(n)$. The
partition function $P(n)$ may be most easily expressed in terms of its
generating function as
\begin{equation}
\displaystyle\sum_{n=0}^{\infty}P(n)x^n=\prod_{k=1}^{\infty}\frac{1}{1-x^k}
=(1+x+x^2+x^3+\cdots)(1+x^2+x^4+x^6+\cdots)(1+x^3+x^6+x^9+\cdots)\cdots
\label{old generate}
\end{equation}
The partition function is interesting for us because it tells us the
number of conjugacy classes of permutations that exist (i.e., have
the same form up to global relabelling).

Since, in our graphs, one of the ports corresponds to a special object
(the port ultimately entangled with Alice's system zero) we must
distinguish our conjugacy classes depending on where within the
permutation this special object sits. To count the number of permutations
with a distinct form (taking into account this special object) we
must cosider a variation of the conventional integer partition theory.
For example, when $n=3$, the partition $2+1$ would become two
partitions when one takes into account that our special object may
sit either in the ``1'' or the ``2''. By contrast, the partition
$1+1+1$ remains a single partition, since although only one of the
``1''s contain our special object the ordering in the sum makes no
difference. As already mentioned every partition corresponds to a
distinct permutation conjugacy class. The number of these classes
is then given the integer partition function, $W(n)$, for our variation
of this problem. Note, that the existence of a special object can only
increase the number of partitions, so we have the trivial lower bound
$P(n)\le W(n)$.

If we think about temporarily labelling our special object as $y$
then it is not hard to see that $W(n)=\frac{\mathrm d}
{\mathrm d y}\big( C_n(y)\big)|_{y=1}$, where
\begin{equation}
\displaystyle\sum_{n=0}^{\infty}C_n(y)x^n=[1+y(x+x^2+x^3+\cdots)]
[1+y(x^2+x^4+x^6+\cdots)][1+y(x^3+x^6+x^9+\cdots)]\cdots
\label{new generate}
\end{equation}
Expanding this out allows us to express our generating function
directly as a function of $x$ by
\begin{equation}
\displaystyle\sum_{n=0}^{\infty}W(n)x^n=x\prod_{k=1}^{\infty}
\frac{1}{1-x^k}+ x^2\prod_{k=1}^{\infty}\frac{1}{1-x^k}
+x^3\prod_{k=1}^{\infty}\frac{1}{1-x^k}+\cdots
= \frac{x}{1-x}\,\prod_{k=1}^{\infty}\frac{1}{1-x^k} \\
\label{real generate}
\end{equation}

We finally note that by dividing both sides of the Eq.~(\ref{real
generate}) by $x$, and replace the right-hand-side of the result by
Eq.~(\ref{old generate}), we may obtain
\begin{equation}
\displaystyle\sum_{n=0}^{\infty}W(n)x^{n-1}=
\frac{1}{1-x}\,\prod_{k=1}^{\infty}
\frac{1}{1-x^k}=\frac{1}{1-x}\,\displaystyle\sum_{n=0}^{\infty}P(n)x^n,
\label{final equation}
\end{equation}
which proves that $W(n)=\sum_{i=0}^{n-1}P(i)$. From the generating
function of $W(n)$ we may also obtain an upper bound for our partition
function as $W(n)\le 1+n\,P(n-1)/2$. We prove this in Supplementary
Derivation S4.

\section*{Acknowledgements}

Zhiwei Wang is grateful to Yi-An Yao and Guo-Mo Zeng for their
suggestions during the research. Zhiwei Wang was supported by the office
of undergraduate education in Jilin University.

\section*{Author contributions statement}

ZWW performed all calulcations and wrote the first draft of the
manuscript. SLB thought up the approach helped find errors during the
analysis and edited the final form of the manuscript. All authors
reviewed the manuscript.

\section*{Additional information}

The authors declare no competing financial interests.

\end{document}